\documentclass[conference]{IEEEtran}
\IEEEoverridecommandlockouts

\usepackage{cite}
\usepackage{amsmath,amssymb,amsfonts}
\usepackage{algorithmic}
\usepackage{graphicx}
\usepackage{textcomp}
\usepackage{xcolor}
\usepackage{hyperref}
\usepackage{booktabs}
\hypersetup{
    colorlinks=true,    
    linkcolor=green,     
    citecolor=green,     
    urlcolor=green       
}\usepackage{multirow}
\usepackage{graphicx}   
\usepackage{subfig}     
\usepackage{colortbl,xcolor}
\usepackage{diagbox}
\usepackage{graphicx}
\usepackage{caption}

\usepackage{pifont}
\usepackage{soul, xcolor, graphicx}

\newif\ifshowcomments
\showcommentstrue 

\ifshowcomments
  
  \definecolor{myblue}{RGB}{200,220,235}
\newcommand{\as}[1]{\sethlcolor{myblue}\hl{[Austin: #1]}}
\else
  \newcommand{\as}[1]{}
  
\fi

\def\BibTeX{{\rm B\kern-.05em{\sc i\kern-.025em b}\kern-.08em
    T\kern-.1667em\lower.7ex\hbox{E}\kern-.125emX}}
\begin{document}

\title{Towards Robust Uncertainty-Aware Speaker Modeling \\
}
\author{
\IEEEauthorblockN{Junjie Li$^{1}$, Yang Xiao$^{2}$, Kong Aik Lee$^{1}$} \IEEEauthorblockA{$^{1}$Department of Electrical and Electronic Engineering, The Hong Kong Polytechnic University,
Hong Kong SAR \\ $^{2}$ The University of Melbourne, Australia
}
}

\maketitle

\begin{abstract}
Speaker embeddings aggregate frame-level acoustic features into compact representations for speaker recognition. Recent uncertainty-aware speaker modeling approaches further characterize the reliability of speaker embeddings by estimating their associated uncertainty. However, existing methods often suffer from inaccurate uncertainty estimation and uncertainty miscalibration under domain shifts. To address these challenges, we propose a robust uncertainty modeling framework from both estimation and adaptation perspectives. Specifically, we introduce an Inter- and Intra-Speaker-Aware Uncertainty Softmax that incorporates both inter-speaker separability and intra-speaker variability into uncertainty learning, enabling uncertainty estimates to better capture the reliability of speaker embeddings. Furthermore, we propose an Uncertainty-Calibrated Domain Adaptation (UCDA) framework to mitigate uncertainty miscalibration caused by domain mismatch. Extensive experiments on both in-domain and cross-domain benchmarks demonstrate that the proposed approach consistently improves uncertainty reliability and speaker recognition robustness.

\end{abstract}

\section{Introduction}
Speaker recognition is widely used in biometric authentication \cite{markowitz2000voice}, personalized human–machine interaction \cite{lin2022personalized}, and intelligent surveillance \cite{kiktova2015speaker}. Modern systems typically extract fixed-dimensional speaker embeddings from variable-length utterances for similarity-based scoring \cite{wang2024overview}. However, in real-world scenarios, speech is frequently corrupted by background noise, reverberation, and channel mismatch \cite{wang2024overview}, introducing severe frame-level uncertainty. Conventional pooling methods, such as average pooling \cite{li2017deep,snyder2018x,wang2021revisiting} and attention-based pooling \cite{okabe2018attentive,zhu2018self,india2019self,zhao2022multi,desplanques20_interspeech,wu2020vector}, rely on deterministic weighting strategies and fail to account for frame reliability, resulting in degraded embedding quality under unconstrained conditions.

To mitigate this, uncertainty modeling represents embeddings as Gaussian distributions \cite{chang2020data,ou2021sdd,shi2019probabilistic,ji2023map,meng2021magface,chen2022fast,li2025xi+,lee2021xi,li2026u3xipushingboundariesspeaker,liu2023disentangling,wang2023incorporating,wang2024cosine,wang22r_interspeech,barahona2025analysis}, where the mean encodes speaker identity and the covariance captures estimation uncertainty to down-weight unreliable frames during pooling and scoring. Recently, the uncertainty-aware additive angular margin softmax (UAAM-Softmax) loss \cite{li2026u3xipushingboundariesspeaker} was introduced to jointly optimize speaker discrimination and uncertainty estimation. Although effective, UAAM-Softmax relies primarily on inter-speaker separability as a supervisory signal, while the intrinsic intra-speaker variability of speaker embeddings is not explicitly considered. As a result, the learned uncertainty estimates may not fully reflect the underlying variability and reliability of speaker embeddings. Moreover, uncertainty estimation is highly sensitive to domain mismatch. Acoustic and environmental variations across datasets induce substantial distribution shifts \cite{li2026u3xipushingboundariesspeaker,wang2024overview}, which can lead to uncertainty miscalibration, where the estimated uncertainty is no longer well aligned with the actual reliability of speaker embeddings, thereby degrading cross-domain performance \cite{ovadia2019can}.

To address the above limitations, we propose a unified framework for robust uncertainty modeling in speaker recognition.
First, we propose an \textbf{Inter- and Intra-Speaker-Aware Uncertainty Softmax} that incorporates both inter-speaker relationships and intra-speaker variability into uncertainty learning. By exploiting complementary supervisory signals, the proposed objective enables uncertainty estimates to better reflect the underlying variability and reliability of speaker embeddings.
Second, we introduce an \textbf{Uncertainty-Calibrated Domain Adaptation (UCDA)} framework that improves the robustness of uncertainty estimation under domain shift. UCDA performs lightweight, label-free adaptation by updating only the uncertainty estimation module and encouraging target-domain uncertainty distributions to move toward a source-domain prior. This targeted calibration improves uncertainty reliability while preserving speaker-discriminative information, resulting in more robust cross-domain speaker recognition.

 

\section{Background: Uncertainty-aware Model}
Recent uncertainty-aware speaker recognition methods \cite{lee2021xi, li2025xi+, li2026u3xipushingboundariesspeaker} model speaker representations as Gaussian distributions to enable reliability-aware feature aggregation. Given frame-level features $\{\mathbf{z}_t\}_{t=1}^{T}$ and predicted diagonal precision matrices $\{\mathbf{L}_t\}_{t=1}^{T}$ for an utterance $X$, the linear-Gaussian formulation \cite{lee2021xi} models each frame as:
\begin{equation}
\mathbf{z}_t = \mathbf{h}+\boldsymbol{\epsilon}_t,
\end{equation}
where $\mathbf{h}$ is the latent speaker variable and $\boldsymbol{\epsilon}_t \sim \mathcal{N}(\mathbf{0},\mathbf{L}_t^{-1})$ denotes frame uncertainty. The accumulated posterior distribution $p(\mathbf{h}|\mathbf{z}_{1:T}) = \mathcal{N}(\mathbf{h}|\boldsymbol{\phi},\mathbf{L}^{-1})$ yields the aggregated mean $\boldsymbol{\phi}$ and residual covariance $\mathbf{L}^{-1}$:
\begin{equation}
\boldsymbol{\phi} = \frac{\sum_{t=1}^{T}\mathbf{L}_t\mathbf{z}_t+\mathbf{L}_p\mathbf{z}_p}{\sum_{t=1}^{T}\mathbf{L}_t+\mathbf{L}_p}, \quad
\mathbf{L}^{-1} = \left(\sum_{t=1}^{T}\mathbf{L}_t+\mathbf{L}_p\right)^{-1}.
\end{equation}

\begin{figure}[tbp]
    \centering
    \includegraphics[width=1\linewidth]{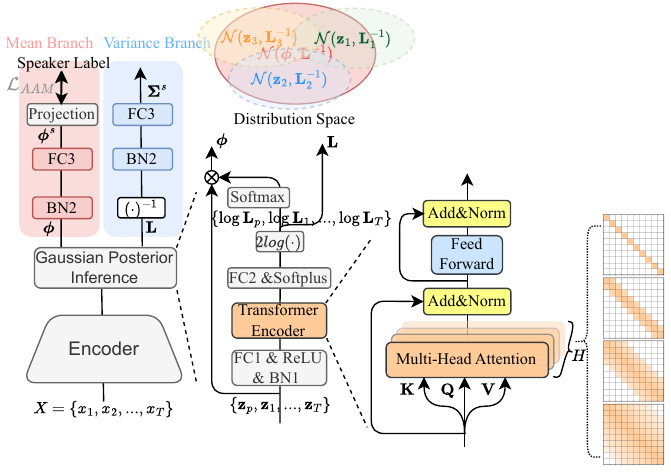}
    \caption{Architecture of the uncertainty-aware speaker model.}
    \label{fig:xi}
    \vspace{-5mm}
\end{figure}

As illustrated in Fig.~\ref{fig:xi}, to obtain the final speaker embedding $\boldsymbol{\phi}^{\text{s}}$ and its coressponding uncertainty $\mathbf{\Sigma}^{\text{s}}$, these statistics are propagated through shared Batch Normalization (BN) and Fully Connected (FC) layers, applying distinct transformations to the mean and variance branches \cite{wang2023incorporating, chen2024pseudo, li2025xi+, li2026u3xipushingboundariesspeaker}. 

Here, we assume a \textbf{diagonal} covariance structure for $\mathbf{L}^{-1}$ and $\mathbf{\Sigma}^{\text{s}}$, allowing element-wise operations across dimensions:
\begin{equation}
\boldsymbol{\phi}^{\text{s}} =
\left(
\frac{\boldsymbol{\phi}-\boldsymbol{\mu}_{\text{bn}}}
{\sqrt{\boldsymbol{\sigma}_{\text{bn}}+\epsilon \mathbf{I}}}
\otimes \boldsymbol{\gamma}_{\text{bn}} + \boldsymbol{\beta}_{\text{bn}}
\right)
\mathbf{A}_{\text{fc}}^\top + \mathbf{b}_{\text{fc}},
\end{equation}

\begin{equation}
\mathbf{\Sigma}^{\text{s}} =
\mathbf{A}_{\text{fc}}
\frac{\mathbf{L}^{-1} \otimes \boldsymbol{\gamma}_{\text{bn}}^2}
{\boldsymbol{\sigma}_{\text{bn}}+\epsilon \mathbf{I}}
\mathbf{A}_{\text{fc}}^\top,
\end{equation}
where $\boldsymbol{\mu}_{\text{bn}}, \boldsymbol{\sigma}_{\text{bn}}, \boldsymbol{\gamma}_{\text{bn}}, \boldsymbol{\beta}_{\text{bn}}$ are BN parameters, and $\mathbf{A}_{\text{fc}}, \mathbf{b}_{\text{fc}}$ denote FC weights and biases.

\section{Uncertainty-aware Softmax}
Conventional Softmax-based objectives, such as additive angular margin Softmax (AAM-Softmax) \cite{deng2019arcface}, operate solely on the point embedding $\boldsymbol{\phi}^{\text{s}}$ and ignore the associated covariance $\boldsymbol{\Sigma}^{\text{s}}$, thereby discarding valuable information about embedding uncertainty.

To address this limitation, $\mathcal{U}^3$-xi \cite{li2026u3xipushingboundariesspeaker} firstly 
 incorporates uncertainty information explicitly into the Softmax formulation, which is called uncertianty-aware AAM (UAAM):
\begin{align}
\mathcal{L}_{\text{UAAM}}
= -\frac{1}{N} \sum_{i=1}^N
\log
\frac{
e^{s \cdot s_u \cos (\theta_{y_i} + m)}
}{
e^{s \cdot s_u \cos (\theta_{y_i} + m)}
+ \sum_{j=1, j\neq y_i}^{C} e^{s \cdot s_u \cos \theta_j}
}, 
\label{eq:UAAM}
\end{align}
where $N$ denotes the batch size, $C$ is the number of speaker classes, $s$ and $m$ are the scale and angular margin hyperparameters, respectively, and $\theta_j$ represents the angle between the normalized embedding and the $j$-th class prototype. $s_u$ denotes the uncertainty-aware scale which incorporates uncertainty information. 

The proposed $\mathcal{L}_{\text{UAAM}}$ introduces an uncertainty-aware scale factor $s_u$, providing explicit supervision for uncertainty estimation. 
The factor $s_u$ is defined as:
\begin{align}
s_u &= \frac{||\boldsymbol{\phi}^\text{s}||}{\sqrt{(\boldsymbol{\phi}^\text{s})^\top (\mathbf{\Lambda} + \mathbf{\Sigma}^\text{s})\boldsymbol{\phi}^\text{s}}}, \label{eq:su} \\
& \propto \frac{\sqrt{(\boldsymbol{\phi}^\text{s})^\top (\mathbf{\Lambda} + \mathbf{\Sigma}^\text{s})^{-1}\boldsymbol{\phi}^\text{s}}}{\|\boldsymbol{\phi}^\text{s}\|}. 
\end{align}
The formulation of $s_u$ is motivated by the Mahalanobis distance, which measures feature confidence under the covariance space. Ideally, uncertainty-aware confidence estimation can be modeled using the inverse covariance matrix $(\mathbf{\Lambda} + \mathbf{\Sigma}^{\text{s}})^{-1}$. However, directly computing the matrix inverse is computationally expensive and often numerically unstable in high-dimensional embedding spaces.
Therefore, we adopt the alternative formulation in (\ref{eq:su}), which avoids explicit matrix inversion while preserving the same inverse relationship between confidence and uncertainty. Specifically, larger uncertainty $\boldsymbol{\Sigma}^{\text{s}}$ leads to a smaller scale factor $s_u$, whereas lower uncertainty results in a larger $s_u$ \cite{li2026u3xipushingboundariesspeaker, 10302546, 10.1609/aaai.v37i12.26760}. In this sense, $s_u$ can be viewed as an efficient approximation of Mahalanobis-style confidence modeling.

 \begin{figure*}[htbp]
     \centering
     \includegraphics[width=1\linewidth]{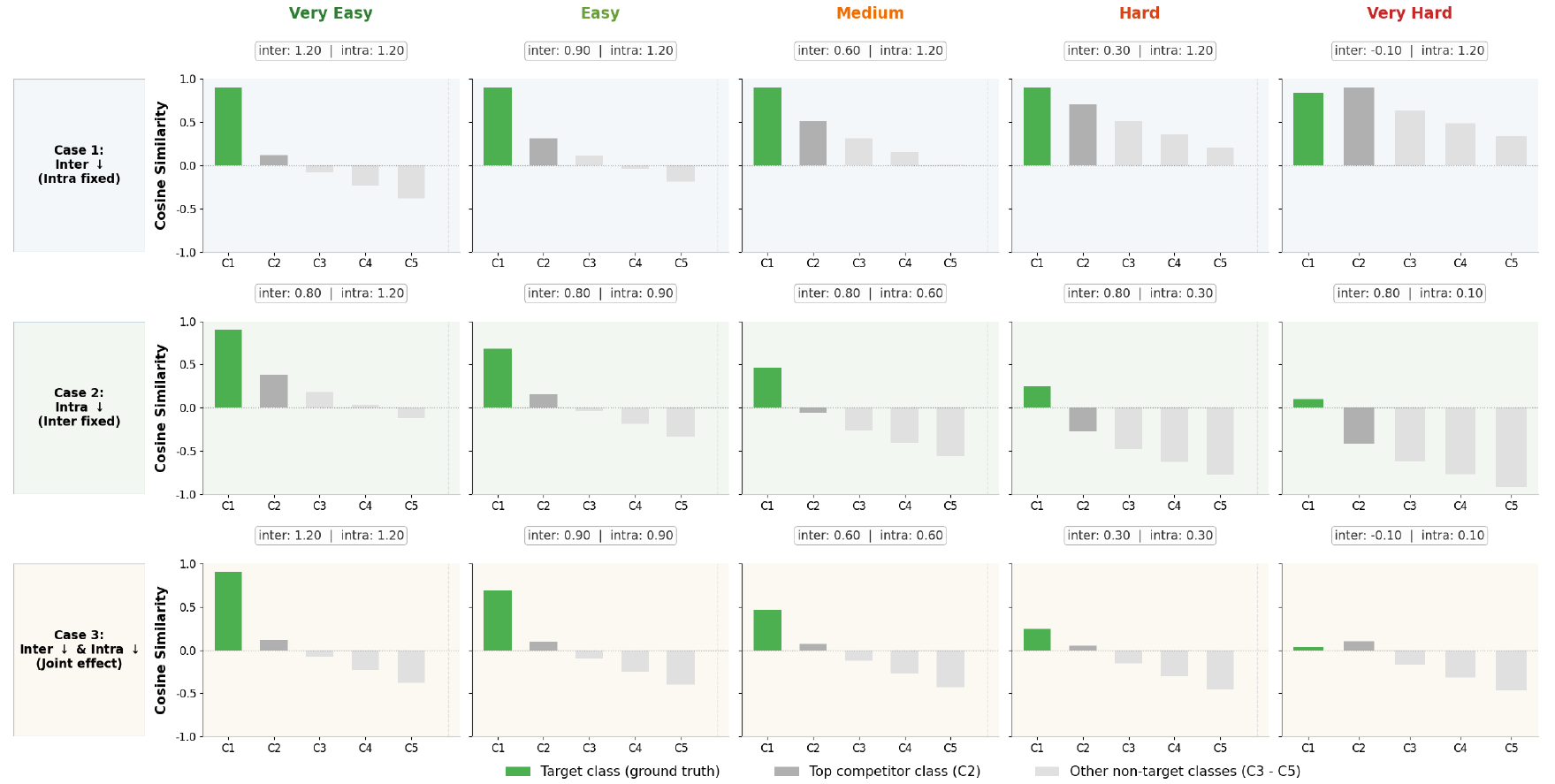}
     \caption{Cosine similarity distributions between sample embeddings and class prototypes across different classification difficulty levels. Each column represents a difficulty setting with progressively reduced inter-speaker separability and/or varying intra-speaker compactness. The y-axis denotes cosine similarity between the input embedding and class prototypes (C1–C5). Green bars indicate the ground-truth class, dark gray the most competitive non-target class, and light gray the remaining classes.}
     \label{fig:inter}
     \vspace{-5mm}
 \end{figure*}

\subsection{Inter-Speaker-Aware Uncertainty Softmax}

To facilitate the learning of $\mathbf{\Sigma}^\text{s}$, the bias term $\mathbf{\Lambda}$ in ~(\ref{eq:su}) should be both positive and data-dependent. At the early stage of training, the predicted uncertainty $\mathbf{\Sigma}^\text{s}$ is usually inaccurate and tends to have very small values \cite{li2026u3xipushingboundariesspeaker}. Consequently, $\mathbf{\Lambda}$ becomes the dominant factor controlling the sample-dependent scale $s_u$. This allows different samples to produce different adaptive scales during optimization.

Based on this motivation, we define $\mathbf{\Lambda}$ using the cosine similarity gap:
\begin{equation}
\boldsymbol{\Lambda} = \big(\lambda - \Lambda_i.\texttt{detach()}\big)\mathbf{I}, \quad \text{\textcircled{1}}
\label{eq:inter}
\end{equation}
where $\lambda$ is a positive scalar hyperparameter ensuring that $\boldsymbol{\Lambda}$ remains positive-definite, and $\mathbf{I}$ denotes the identity matrix. The $\texttt{detach()}$ operation prevents $\Lambda_i$ from receiving gradient updates during training. The term $\Lambda_i$ measures the inter-speaker hardness:
\begin{equation}
\Lambda_i = \cos\theta_{y_i} - \max_{j \neq y_i} \cos\theta_j,
\end{equation}
where $\cos\theta_{y_i}$ denotes the cosine similarity between the embedding and the target class prototype, and $\max_{j \neq y_i} \cos\theta_j$ represents the similarity to the most competitive non-target class prototype. Larger $\Lambda_i$ values indicate easier samples with
better inter-speaker separability. 

However, modulating $s_u$ solely through $\boldsymbol{\Lambda}$ may restrict the expressiveness and flexibility of the uncertainty-aware scaling mechanism. To further incorporate sample hardness into the scaling process, we directly introduce the hardness-aware factor into the scale formulation:
\begin{equation}
s_u = \exp(\Lambda_i) \cdot 
\frac{||\boldsymbol{\phi}^\text{s}||}
{\sqrt{(\boldsymbol{\phi}^\text{s})^\top (\boldsymbol{\Lambda} + \boldsymbol{\Sigma}^\text{s})\boldsymbol{\phi}^\text{s}}}.
\quad \text{\textcircled{2}}
\label{eq:inter_scale}
\end{equation}
The exponential term amplifies the distinction
between easy and hard samples and enhances the
sensitivity of uncertainty-aware scaling.

\subsection{Inter- and Intra-Speaker-Aware Uncertainty Softmax}

However, the formulation in (\ref{eq:inter}) primarily captures inter-speaker separability, i.e., the margin between the target class and the most competitive non-target class, while neglecting intra-speaker compactness. In practice, the difficulty of a sample is jointly determined by both its separation from competing classes (inter-speaker hardness) and its alignment with the most similar class prototype across all categories (intra-speaker hardness) \cite{liu2025adaspeaker}. Consequently, samples with similar inter-speaker margins may still exhibit different reliability due to variations in their best-prototype similarity.

Fig.~\ref{fig:inter} illustrates this distinction. inter-speaker hardness reflects how close the target class is to its strongest competitor, whereas intra-speaker hardness reflects how strongly the sample aligns with the most similar prototype among all classes. As shown in Fig.~\ref{fig:inter}, both factors vary independently or jointly under different settings, leading to different classification difficulty levels.

Motivated by this analysis, we propose a variation of uncertainty-aware scaling:
\begin{align}
\boldsymbol{\Lambda} = \big(\lambda  - \big(\Lambda_i \cdot \Lambda_j\big).\texttt{detach()}\big)\mathbf{I}, \quad \text{\textcircled{3}}
\label{eq:intra}
\end{align}
where the intra-speaker confidence term is defined as:
\begin{align}
\Lambda_j = \exp\left( \max_{j \in \{1,\dots,C\}} \cos\theta_j \right),
\end{align}
with $\cos\theta_j$ denoting the cosine similarity between the embedding and the $j$-th class prototype. The exponential function is adopted to ensure that the intra term remains strictly positive and preserves a consistent scaling direction during optimization. In addition, $\exp(\cdot)$ enlarges the relative difference between highly confident and ambiguous samples, making the uncertainty modulation more sensitive to variations in prototype alignment.

Similar to (\ref{eq:inter_scale}), we further incorporate the joint inter- and intra-speaker-aware hardness factor directly into the uncertainty-aware scaling function:
\begin{equation}
s_u = \exp (\Lambda_i \cdot \Lambda_j) \cdot
\frac{||\boldsymbol{\phi}^\text{s}||}
{\sqrt{(\boldsymbol{\phi}^\text{s})^\top (\boldsymbol{\Lambda} + \boldsymbol{\Sigma}^\text{s})\boldsymbol{\phi}^\text{s}}},
\quad \text{\textcircled{4}}
\label{eq:inter_intra}
\end{equation}where the exponential modulation jointly considers inter-speaker separability and intra-speaker compactness, enabling a more discriminative and adaptive uncertainty-aware scaling behavior.

\begin{table*}[htbp]
\centering
\setlength{\aboverulesep}{1pt}
\setlength{\belowrulesep}{1pt}
\caption{Overall results on Voxceleb1 in terms of the EER and minDCF. The best performance is shown in \textbf{bold}. Results in gray denote we apply uncertainty-aware cosine score \cite{li2026u3xipushingboundariesspeaker,li2025xi+}. RI denotes relative improvements. }
\resizebox{\linewidth}{!}{
\begin{tabular}{c | c | c| c  |  c c | c c | c c | c | cc | c}
\toprule
\multirow{3}{*}{Exp.} & \multirow{3}{*}{Model} & \multirow{3}{*}{\# Param.} &  \multirow{3}{*}{Loss} 
& \multicolumn{7}{c|}{In-domain} 
& \multicolumn{3}{c}{Cross-domain} \\
\cmidrule{5-14}
& & & &
\multicolumn{2}{c|}{Vox1-O} & 
\multicolumn{2}{c|}{Vox1-E} & 
\multicolumn{2}{c|}{Vox1-H} & 
RI (\%) & 
\multicolumn{2}{c|}{CNCeleb} & 
RI (\%) \\
\cmidrule{5-14}
& & & &
EER & minDCF & 
EER & minDCF & 
EER & minDCF & 
 & 
EER & minDCF & 
 \\
\midrule
1 & ECAPA512 \cite{wang2023wespeaker}   & 6.19 M & AAM-Softmax \cite{deng2019arcface} &   
1.069 & 0.122 & 1.209 & 0.136 & 2.310 & 0.226 & Benchmark & 15.314 & 0.633 & Benchmark \\ \midrule 

\multirow{2}{*}{2} & \multirow{2}{*}{ECAPA512+$\mathcal{U}^3$-xi} & \multirow{2}{*}{6.69 M} 
& \multirow{2}{*}{UAAM-Softmax \textcircled{1} \cite{li2026u3xipushingboundariesspeaker}}  
& 0.894 & 0.122 & 1.075 & 0.121 & 2.006 & 0.199 &10.60  & 13.760 & 0.578 & 9.42 \\  
& & & & \cellcolor{gray!20}0.851 & \cellcolor{gray!20}0.113 & \cellcolor{gray!20}1.035 & \cellcolor{gray!20}0.115 & \cellcolor{gray!20}1.926 & \cellcolor{gray!20}0.191 &  \cellcolor{gray!20}14.95 & \cellcolor{gray!20}10.690 & \cellcolor{gray!20}0.877 &  \cellcolor{gray!20}-4.19\\ \midrule 

\multirow{2}{*}{3} & \multirow{2}{*}{ECAPA512+$\mathcal{U}^3$-xi}
& \multirow{2}{*}{6.69 M} & \multirow{2}{*}{UAAM-Softmax \textcircled{3}}  
& 0.930 & 0.117 & 1.102 & 0.127 & 2.114 & 0.206 & 8.32 & 14.545 & 0.642 & 1.80 \\  
& & & & \cellcolor{gray!20}0.920 & \cellcolor{gray!20}0.117 & \cellcolor{gray!20}1.094 & \cellcolor{gray!20}0.125 & \cellcolor{gray!20}2.093 & \cellcolor{gray!20}0.205 &\cellcolor{gray!20} 9.05 & \cellcolor{gray!20}12.763 & \cellcolor{gray!20}0.873 & \cellcolor{gray!20}-10.65 \\ \midrule 

\multirow{2}{*}{4} & \multirow{2}{*}{ECAPA512+$\mathcal{U}^3$-xi}
& \multirow{2}{*}{6.69 M} & \multirow{2}{*}{UAAM-Softmax \textcircled{1}+\textcircled{2}}  
& 0.957 & 0.122 & 1.069 & 0.121 & 2.036 & 0.202 & 9.26 & 12.920 & \textbf{0.564} &  13.27\\  
& & & & \cellcolor{gray!20}\textbf{0.819} & \cellcolor{gray!20}0.100 & \cellcolor{gray!20}0.988 & \cellcolor{gray!20}0.113 & \cellcolor{gray!20}1.879 & \cellcolor{gray!20}0.190 &\cellcolor{gray!20} 18.53 & \cellcolor{gray!20}\textbf{9.237} & \cellcolor{gray!20}1.000 & \cellcolor{gray!20}-9.15  \\ \midrule 

\multirow{2}{*}{5} & \multirow{2}{*}{ECAPA512+$\mathcal{U}^3$-xi}
& \multirow{2}{*}{6.69 M} & \multirow{2}{*}{UAAM-Softmax \textcircled{1}+\textcircled{4}}  
& 0.936 & 0.102 & 1.050 & 0.122 & 1.978 & 0.195 & 13.40 & 13.974 & 0.581 & 8.48 \\  
& & & & \cellcolor{gray!20}0.840 & \cellcolor{gray!20}\textbf{0.086} & \cellcolor{gray!20}\textbf{0.965} & \cellcolor{gray!20}\textbf{0.110} & \cellcolor{gray!20}\textbf{1.833} & \cellcolor{gray!20}\textbf{0.189} & \cellcolor{gray!20} 21.22 & \cellcolor{gray!20}10.781 & \cellcolor{gray!20}0.835 & \cellcolor{gray!20}-1.16 \\ \midrule  \midrule

6 & ECAPA512   & 6.19 M & AM-Softmax \cite{wang2018additive,wang2018cosface} & 
1.005 & 0.107 & 1.206 & 0.133 & 2.254 & 0.221 & Benchmark & 14.162 & 0.611 & Benchmark \\ \midrule 

\multirow{2}{*}{7} & \multirow{2}{*}{ECAPA512+$\mathcal{U}^3$-xi}  
& \multirow{2}{*}{6.69 M} & \multirow{2}{*}{UAM-Softmax \textcircled{1}+\textcircled{4}} 
& 0.888 & 0.099 & 1.076 & 0.119 & 1.973 & 0.186 &11.46  & 12.436 & \textbf{0.553} & 10.84 \\ 
& & & & \cellcolor{gray!20}\textbf{0.808} & \cellcolor{gray!20}\textbf{0.084} & \cellcolor{gray!20}\textbf{0.991} & \cellcolor{gray!20}\textbf{0.109} & \cellcolor{gray!20}\textbf{1.794} & \cellcolor{gray!20}\textbf{0.178} & \cellcolor{gray!20}19.46 & \cellcolor{gray!20}\textbf{9.411} & \cellcolor{gray!20}1.000 & \cellcolor{gray!20}-15.03 \\ \midrule  \midrule

8 & ECAPA512 & 6.19 M & SphereFace2 \cite{han2023exploring, wen2021sphereface2}  
& 0.963 & 0.108 & 1.121 & 0.125 & 1.967 & 0.199 & Benchmark & 12.582 & 0.573 & Benchmark \\ \midrule 

\multirow{2}{*}{9} & \multirow{2}{*}{ECAPA512+$\mathcal{U}^3$-xi} 
& \multirow{2}{*}{6.69 M} & \multirow{2}{*}{USphereFace2 \textcircled{1} + \textcircled{4}} 
&  0.856&0.104  &1.035  &0.119  & 1.918 &  0.196& 5.21 &12.265  &\textbf{0.550}  & 3.27 \\ 
  &  &  &  & \cellcolor{gray!20}\textbf{0.739}&\cellcolor{gray!20} \textbf{0.102}&\cellcolor{gray!20}\textbf{0.965} &\cellcolor{gray!20}\textbf{0.108} &\cellcolor{gray!20}\textbf{1.771}  & \cellcolor{gray!20}\textbf{0.178} & \cellcolor{gray!20}12.81 & \cellcolor{gray!20}\textbf{10.560}& \cellcolor{gray!20}0.624  &\cellcolor{gray!20}3.59\\ \midrule  \midrule

- & CAM++ \cite{wang2023cam++} & 7.2 M & - & 
0.808 & 0.109 & 0.931 & 0.109 & 1.863 & 0.179 & -- & 15.179 & 0.635 & -- \\ \midrule 

- & Gemini SD-ResNet38 \cite{liu2024golden} & 6.72 M & - & 
1.085 & 0.099 & 1.130 & 0.117 & 1.974 & 0.185 & -- & 11.507 & 0.553 & -- \\ \midrule 

- & ECAPA1024 \cite{desplanques20_interspeech} & 14.65 M & - & 
0.856 & 0.090 & 1.072 & 0.117 & 2.059 & 0.205 & -- & 15.532 & 0.670 & -- \\ \midrule 

- & ResNet34 \cite{zeinali2019but} & 6.63 M & - & 
0.867 & 0.091 & 1.049 & 0.121 & 1.960 & 0.192 & -- & 11.090 & 0.488 & -- \\ \bottomrule 

\end{tabular}}
\label{tab:scale}
\vspace{-5mm}
\end{table*}

\section{Uncertainty-Calibrated Domain Adaptation}

Uncertainty estimation in speaker recognition is highly sensitive to acoustic variations such as noise, reverberation, channel mismatch, and speech duration. As a result, models trained on source-domain data often suffer significant performance degradation under cross-domain conditions \cite{li2026u3xipushingboundariesspeaker}.

This issue arises because uncertainty modeling is implicitly learned from source-domain acoustic statistics, making the estimated uncertainty distributions poorly aligned with unseen target domains. Consequently, the uncertainty estimator produces unreliable confidence estimates under domain shift.

To address this problem, we propose an Uncertainty-Calibrated Domain Adaptation (UCDA) framework, which performs distribution alignment in the uncertainty space for label-free adaptation. As illustrated in Fig.~\ref{fig:uctta}, uncertainty distributions exhibit clear domain shifts across datasets, e.g., between VoxCeleb1 and CNCeleb.

Unlike conventional feature-level domain adaptation methods, such as CORAL \cite{sun2016deep}, MMD-based alignment \cite{long2015learning}, and adversarial approaches like DANN \cite{ganin2016domain}, which primarily focus on aligning embedding distributions \cite{huang2024robust,chen2021self}, UCDA  targets the uncertainty space, providing a complementary perspective for domain adaptation in speaker recognition.

Specifically, UCDA introduces an uncertainty calibration objective that aligns the uncertainty distribution of target-domain utterances with a source-domain prior. We model utterance-level uncertainty vectors using a Gaussian distribution estimated from the source domain, and optimize a negative log-likelihood (NLL) objective \cite{kendall2017uncertainties}:

\begin{align}
\vspace{-2mm}
\mathcal{L}_{\text{UCDA}}
=
\frac{1}{B}
\sum_{i=1}^{B}
-\log
\mathcal{N}
\left(
\boldsymbol{\Sigma}^{\mathrm{tgt}}_{i}
\,\middle|\,
\boldsymbol{\mu}_{\mathrm{src}},
\boldsymbol{\sigma}^{2}_{\mathrm{src}}
\right),
\end{align}
where $\boldsymbol{\mu}_{\mathrm{src}}$ and $\boldsymbol{\sigma}^{2}_{\mathrm{src}}$ are the  Gaussian parameters estimated from the source-domain training set, and $\boldsymbol{\Sigma}^{\mathrm{tgt}}_{i} \in \mathbb{R}^{D}$ denotes the utterance-level uncertainty vector of the $i$-th target-domain sample.

The source-domain statistics are computed as:
\begin{equation}
\boldsymbol{\mu}_{\mathrm{src}}
=
\frac{1}{N}
\sum_{i=1}^{N}
\boldsymbol{\Sigma}^{\mathrm{src}}_{i},
\qquad
\boldsymbol{\sigma}^{2}_{\mathrm{src}}
=
\frac{1}{N}
\sum_{i=1}^{N}
\left(
\boldsymbol{\Sigma}^{\mathrm{src}}_{i}
-
\boldsymbol{\mu}_{\mathrm{src}}
\right)^{2}.
\end{equation}
Importantly, UCDA performs distribution-level alignment between source and target uncertainty using a likelihood-based formulation defined on a fixed source-domain prior. The proposed NLL objective operates on each target utterance independently, without requiring any statistics estimated from the target domain, making it suitable for real-world deployment scenarios where test samples arrive sequentially.
By maximizing the likelihood under the source-domain uncertainty distribution, UCDA encourages the target uncertainty distribution to move toward the source-domain reliability patterns, thereby improving robustness under domain shift.

To ensure stable speaker embedding estimation during domain adaptation, \textbf{all model parameters are frozen except those associated with the uncertainty prediction module, i.e., the Gaussian posterior inference component in the pooling layer} (as illustrated in Fig.~\ref{fig:xi}). By restricting parameter updates exclusively to the uncertainty estimation pathway, the proposed strategy effectively avoids perturbing the learned speaker embedding space and classifier decision boundaries, thereby enabling lightweight, label-free, and fully unsupervised adaptation with strong robustness against catastrophic drift.

\begin{figure*}[htbp]
    \centering
    \subfloat[Exp. 2]{%
        \includegraphics[width=0.24\linewidth]{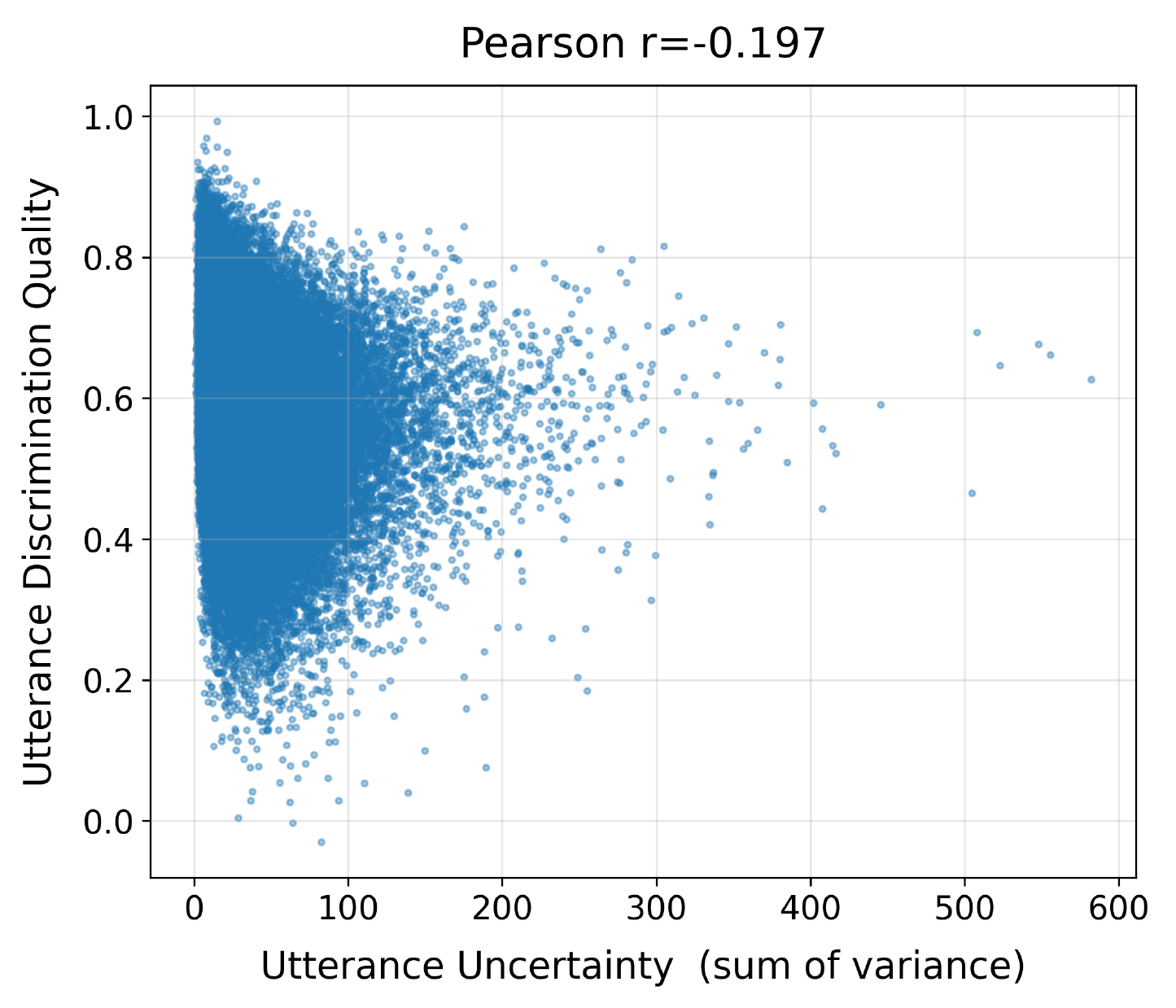}
    }
    \subfloat[Exp. 3]{%
        \includegraphics[width=0.24\linewidth]{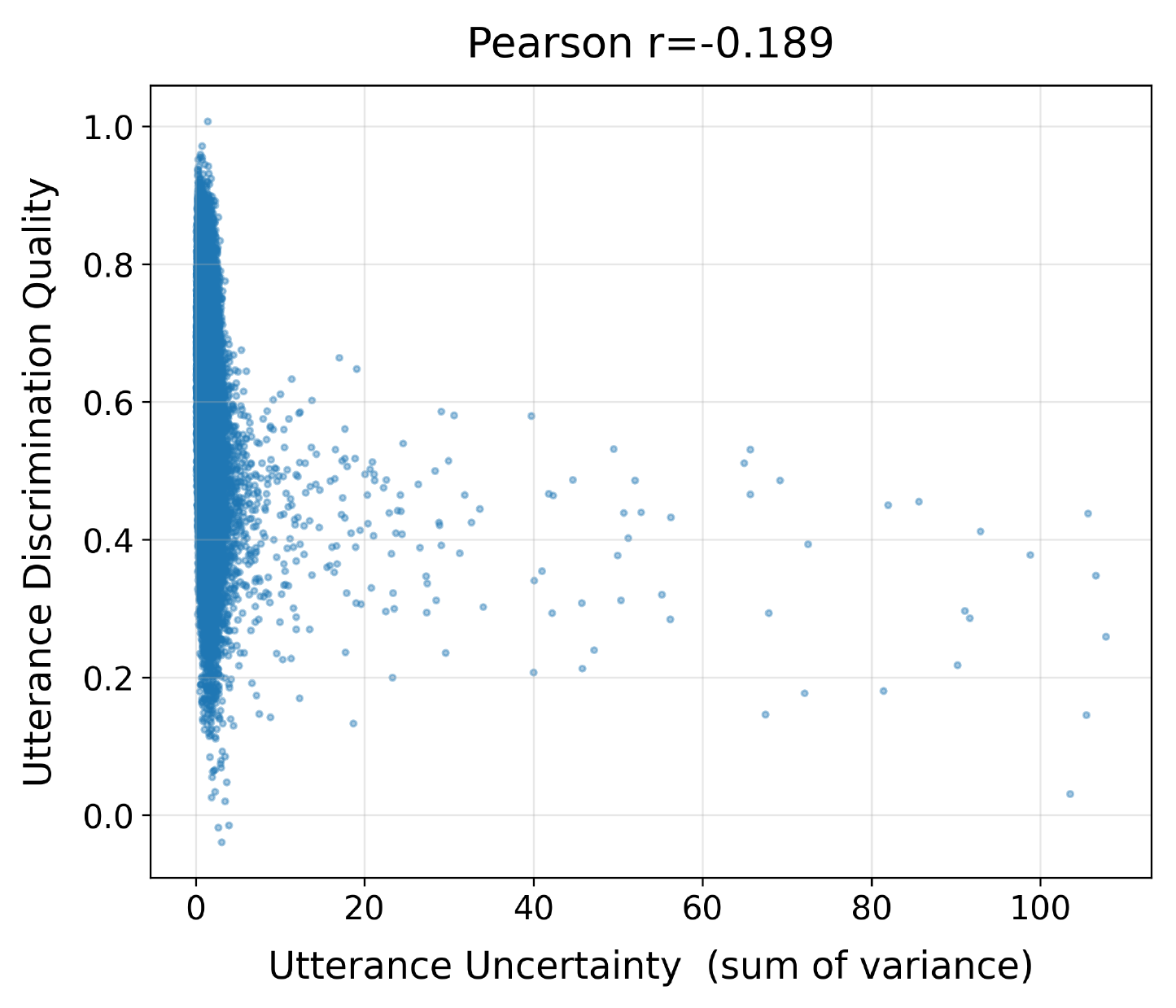}
    }
    \subfloat[Exp. 4]{%
        \includegraphics[width=0.24\linewidth]{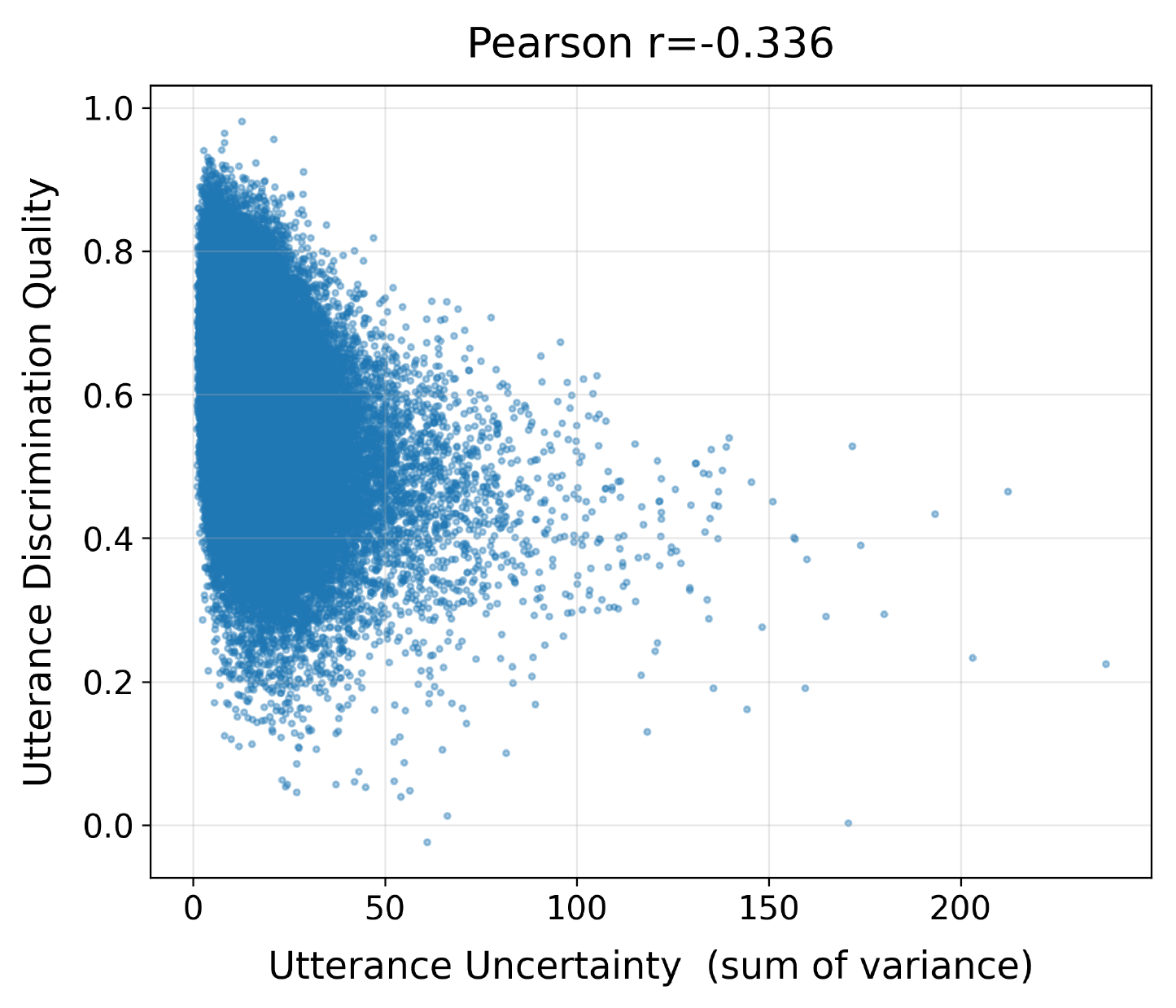}
    }
    \subfloat[Exp. 5]{%
        \includegraphics[width=0.24\linewidth]{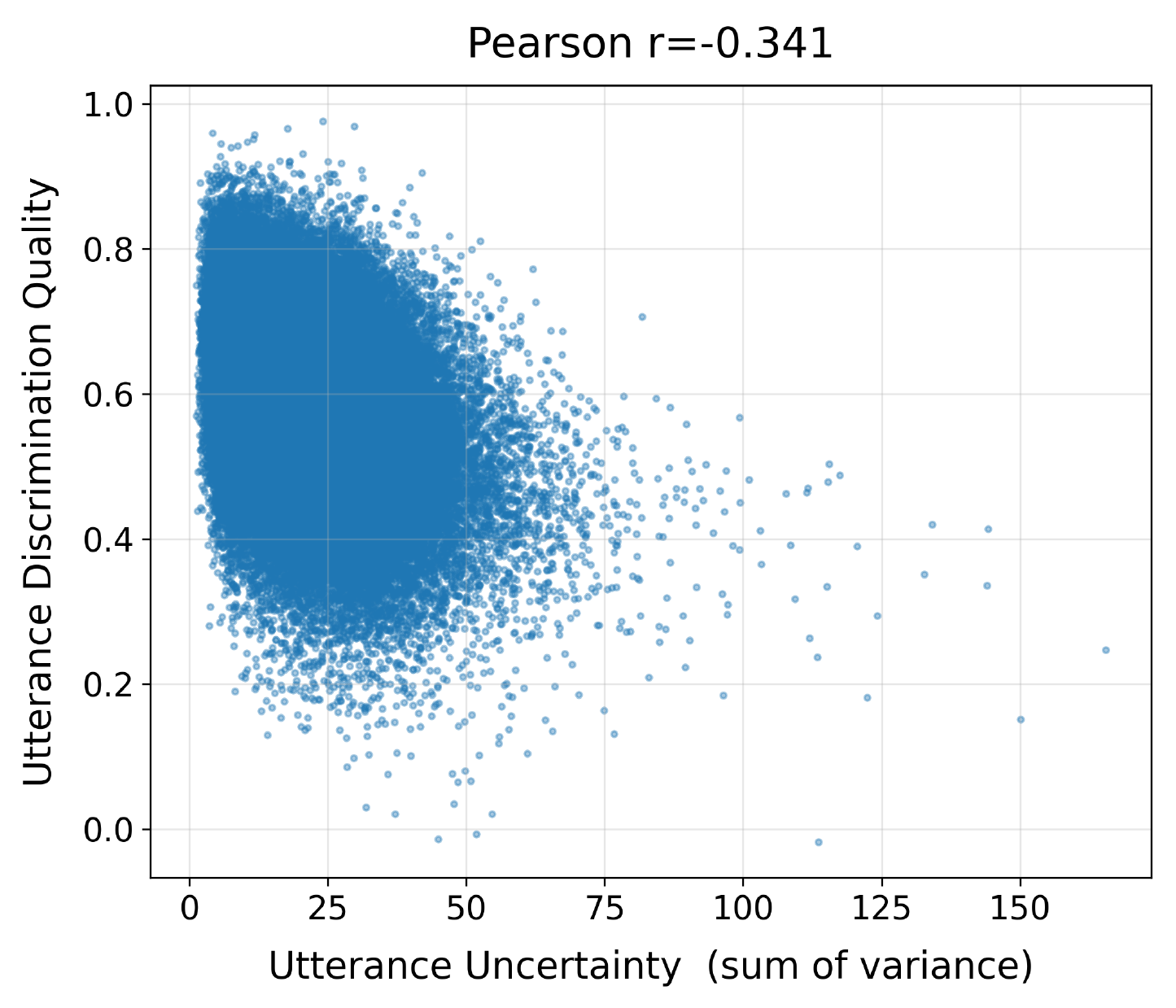}
    }
    \caption{Relationship between utterance-level uncertainty and speaker discrimination quality on VoxCeleb1. Each point corresponds to one utterance.  A strong negative correlation demonstrates that the learned uncertainty reliably reflects the embedding reliability.}
    \label{fig:visual}
    \vspace{-6mm}
\end{figure*}

\section{Experiments and Datasets}

We follow the VoxCeleb v2 training pipeline provided by the WeSpeaker toolkit
 \cite{wang2023wespeaker}, retaining the default hyperparameters. Training is conducted for 150 epochs on 2-second audio segments. 
 The default scale $s$  is set to 32, and the angular margin is gradually increased from 0 to 0.2 between epochs 20 and 40, after which it remains constant. For the final model, parameters from the last 10 checkpoints are averaged.  Data augmentation is applied throughout training, including additive noise from the MUSAN corpus \cite{snyder2015musan}, simulated reverberation using room impulse responses (RIRs) from the RIR database \cite{ko2017study}, and speed perturbation at factors of 0.9×, 1×, and 1.1×. 

All models are trained on VoxCeleb2 \cite{chung2018voxceleb2} and evaluated on the in-domain VoxCeleb1 \cite{nagrani2017voxceleb} benchmark, as well as  cross-domain dataset: CNCeleb \cite{fan2020cn}. 

We report the performances in terms of the equal error rate (EER) and the minimum detection cost function (minDCF) with P$_\text{target}$ = 0.01 and C$_\text{FA}$ = C$_\text{Miss}$ = 1. The scores are produced by calculating the cosine distance between embeddings. 

\section{Results}
\subsection{Effectiveness of Inter- and Intra-Speaker-Aware Uncertainty Modeling}

In this section, we compare the effectiveness of our proposed uncertainty scaling method, as summarized in Table \ref{tab:scale}. Results are evaluated using both the standard cosine score:

\begin{equation}
s_{\text{cos}}(\boldsymbol{\phi}_e^\text{s},\boldsymbol{\phi}_t^\text{s})
= \frac{\langle \boldsymbol{\phi}_e^\text{s},\boldsymbol{\phi}_t^\text{s} \rangle}
{\lVert \boldsymbol{\phi}_e^\text{s} \rVert , \lVert \boldsymbol{\phi}_t^\text{s} \rVert},
\label{eq:cos}
\end{equation}
and the uncertainty-aware cosine score \cite{li2026u3xipushingboundariesspeaker,li2025xi+}:
\begin{equation}
s_\text{ucos}
= \frac{\langle \boldsymbol{\phi}_e^\text{s},\boldsymbol{\phi}_t^\text{s} \rangle}
{\sqrt{(\boldsymbol{\phi}_e^\text{s})^\top (\mathbf{I}+\boldsymbol{\Sigma}_e^\text{s})^{-1}\boldsymbol{\phi}_e^\text{s}}
\sqrt{(\boldsymbol{\phi}_t^\text{s})^\top (\mathbf{I}+\boldsymbol{\Sigma}_t^\text{s})^{-1}\boldsymbol{\phi}_t^\text{s}}},
\label{equ:ucos}
\end{equation}
where  $\boldsymbol{\phi}_e^\text{s}$ and $\boldsymbol{\phi}_t^\text{s}$ denote the enrollment and test embeddings, respectively, and $\boldsymbol{\Sigma}_e^\text{s}$ and $\boldsymbol{\Sigma}_t^\text{s}$
 represent their corresponding uncertainties.

\subsubsection{In-domain Test}
Exp.~1 corresponds to the baseline ECAPA-TDNN system trained with conventional AAM-Softmax, which does not incorporate any uncertainty modeling. Exp. 2 reports the performance of  previous uncertainty-aware framework, $\mathcal{U}^3$-xi trained with UAAM-Softmax \cite{li2026u3xipushingboundariesspeaker}, where only inter-speaker separability is exploited as hardness supervision. Compared to Exp. 1, Exp. 2 achieves consistent improvements across all in-domain test sets, especially after applying uncertainty-aware score, 
demonstrating the effectiveness of uncertainty-aware training in enhancing embedding robustness. 

In Exp. 3, we further incorporate intra-speaker compactness into uncertainty modeling by redefining the bias term $\boldsymbol{\Lambda}$ according to (\ref{eq:intra}). Contrary to expectation, this modification does not provide additional gains over Exp. 2 and even degrades performance in most evaluation conditions.

One possible reason is related to the stabilizing constant $\lambda$ in (\ref{eq:inter}) and (\ref{eq:intra}). Since the numerical ranges of the inter-speaker term and the joint inter- and intra-speaker term differ substantially, different values of $\lambda$ are required to ensure the positive definiteness of $\boldsymbol{\Lambda}$. In our experiments, the smallest feasible values were $\lambda=0.5$ for (\ref{eq:inter}) and $\lambda=1.2$ for (\ref{eq:intra}). The considerably larger value required by the latter reduces the relative contribution of the uncertainty term $\boldsymbol{\Sigma}^{\text{s}}$ in the uncertainty-aware scaling function. As reported in \cite{li2026u3xipushingboundariesspeaker}, excessively large stabilizing constants tend to weaken uncertainty modulation, thereby diminishing the influence of uncertainty information during optimization. Consequently, the potential benefit of incorporating intra-speaker compactness may be partially offset, leading to the suboptimal performance observed in Exp. 3. Consequently, the observed degradation may not necessarily indicate that intra-speaker compactness is uninformative for uncertainty modeling; rather, it suggests that the current formulation introduces an additional optimization trade-off through the stabilizing constant $\lambda$.


To further enhance the expressiveness of uncertainty scaling, Exp. 4 and Exp. 5 explicitly incorporate the inter-speaker term (Eq.~\ref{eq:inter_scale}) and the joint inter- and intra-speaker term (Eq.~\ref{eq:inter_intra}) into the uncertainty-aware scale factor, respectively. Both variants yield consistent improvements across all in-domain test sets.  Among all configurations, Exp. 5 delivers the best overall in-domain performance. These results suggest that directly incorporating both inter-speaker separability and intra-speaker compactness into the uncertainty-aware scaling function provides a more informative measure of sample difficulty, leading to more effective uncertainty modulation.
\subsubsection{Analysis of Learned Uncertainty and Sample Hardness}

Following the findings of $\mathcal{U}^3$-xi \cite{li2026u3xipushingboundariesspeaker}, the learning of uncertainty $\boldsymbol{\Sigma}^{\text{s}}$ is primarily driven by the uncertainty-aware scaling factor $s_u$. Therefore, a more informative scaling function is expected to facilitate the learning of more reliable uncertainty representations.

To examine whether the predicted uncertainty correlates with utterance difficulty, we visualize the relationship between utterance-level uncertainty and speaker discrimination quality in Fig.~\ref{fig:visual}. The utterance-level uncertainty of sample $i$ is computed as:
\begin{equation}
u_i=\sum_{d=1}^{D}\boldsymbol{\Sigma}^{\text{s}}_{i,d},
\end{equation}
while the corresponding discrimination quality is defined as:
\begin{equation}
q_i=
\mathbb{E}_{j\in\mathcal{P}_i}
[\cos(\boldsymbol{e}_i,\boldsymbol{e}_j)]
\mathbb{E}_{k\in\mathcal{N}_i}
[\cos(\boldsymbol{e}_i,\boldsymbol{e}_k)],
\end{equation}
where $\mathcal{P}_i$ and $\mathcal{N}_i$ denote the target and non-target trial sets associated with utterance $i$, respectively.

We further compute the Pearson correlation coefficient between $u_i$ and $q_i$ for each model. The results are consistent with the performance trends reported in Table~\ref{tab:scale}. In particular, Exp.~5 exhibits the strongest negative correlation, indicating that higher predicted uncertainty is more consistently associated with lower discrimination quality. This suggests that the proposed inter- and intra-speaker-aware uncertainty scaling produces uncertainty estimates that better reflect the intrinsic difficulty of speaker embeddings.

\subsubsection{Cross-domain Test}

The performance trends under cross-domain evaluation differ from those observed in the in-domain setting. Under the standard cosine scoring, the proposed models consistently achieve improvements. After applying uncertainty-aware scoring (shown in gray), the EER is further reduced across all systems. However, the minDCF metric degrades in most cases and even reaches $1.000$. 
This discrepancy suggests that the uncertainty estimation becomes less reliable under cross-domain conditions.

In addition, we extend the proposed uncertainty-aware formulation to additive margin Softmax (AM-Softmax) \cite{wang2018additive,wang2018cosface} and SphereFace2 \cite{han2023exploring, wen2021sphereface2}. As shown in Exp.~7 and Exp.~9 of Table~\ref{tab:scale}, both methods benefit from the proposed inter- and intra-speaker-aware  uncertainty scaling, demonstrating its generality across different classification objectives. Notably, SphereFace2 achieves the best overall performance under most evaluation metrics and exhibits improved robustness under cross-domain conditions.

\subsection{The Effectiveness of UCDA}
\begin{table}[t]
\centering
\caption{
Performance on CNCeleb after applying UCDA. Gray cells indicate results obtained with uncertainty-aware cosine scoring. RI denotes the average relative improvement. Experiments 10–12 apply UCDA with different learning rates (LR).
}

\begin{tabular}{c|c|c|cc|c}
\toprule
Exp. & LR & UCDA & EER & minDCF & RI (\%) \\
\midrule

8
& --
& \ding{55}
& 12.582
& 0.573
& Benchmark
\\

\midrule

\multirow{2}{*}{9}
& \multirow{2}{*}{--}
& \multirow{2}{*}{\ding{55}}
& 12.265
& 0.550
& 2.20
\\
&
&
& \cellcolor{gray!20}10.560
& \cellcolor{gray!20}0.624
& \cellcolor{gray!20}3.59
\\

\midrule\midrule

\multirow{2}{*}{10}
& \multirow{2}{*}{$10^{-5}$}
& \multirow{2}{*}{\ding{51}}
& 12.560
& 0.547
& 2.36
\\
&
&
& \cellcolor{gray!20}12.022
& \cellcolor{gray!20}0.528
& \cellcolor{gray!20}6.15
\\

\midrule

\multirow{2}{*}{11}
& \multirow{2}{*}{$10^{-6}$}
& \multirow{2}{*}{\ding{51}}
& 12.508
& 0.542
& 3.00
\\
&
&
& \cellcolor{gray!20}11.811
& \cellcolor{gray!20}0.524
& \cellcolor{gray!20}7.34
\\

\midrule

\multirow{2}{*}{12}
& \multirow{2}{*}{$10^{-7}$}
& \multirow{2}{*}{\ding{51}}
& 12.470
& 0.538
& 3.50
\\
&
&
& \cellcolor{gray!20}11.667
& \cellcolor{gray!20}0.526
& \cellcolor{gray!20}7.74
\\

\bottomrule

\end{tabular}

\vspace{-3mm}
\label{tab:UCTTA}
\end{table}

In this section, we evaluate the effectiveness of the proposed Uncertainty-Calibrated Domain Adaptation (UCDA) framework. The adaptation is conducted on Exp. 9 for 5 epochs.  As shown in Table~\ref{tab:UCTTA}, we investigate different learning rates, all of which consistently bring performance improvements, though with varying magnitudes.

Compared with Exp. 9, UCDA demonstrates a clear trade-off between EER and minDCF. Specifically, while standard cosine scoring yields marginal improvements on EER, it shows limited gains on minDCF. In contrast, uncertainty-aware cosine scoring significantly improves both metrics in most settings, indicating its effectiveness in stabilizing decision boundaries after domain adaptation.

From the perspective of relative improvement (RI), UCDA consistently achieves substantial gains over the baseline, with the best overall performance obtained at a learning rate of $10^{-7}$. This suggests that a smaller learning rate provides a more stable adaptation process, leading to more reliable uncertainty calibration. Overall, the proposed UCDA effectively enhances system robustness, despite minor metric-specific trade-offs.

\begin{figure}[htbp]
\vspace{-3mm}
    \centering
    \includegraphics[width=1\linewidth]{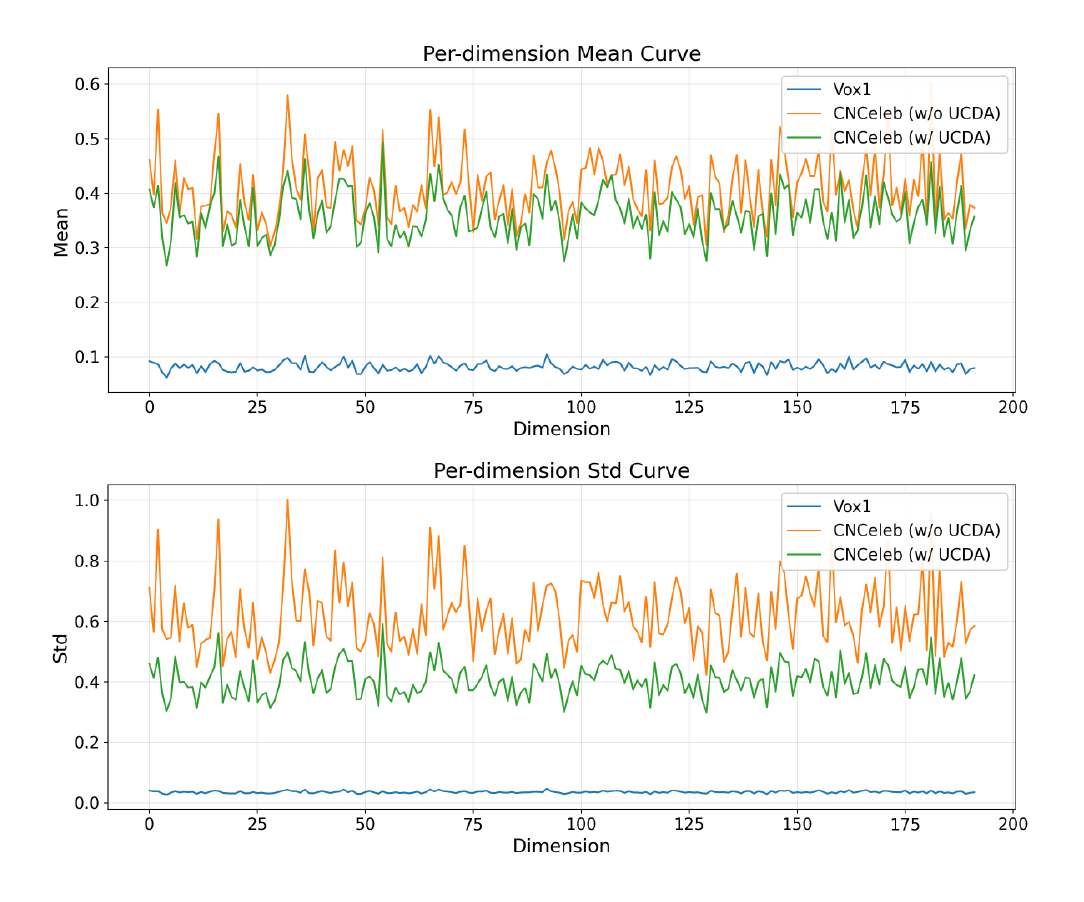}
    \vspace{-10mm}
    \caption{Uncertainty distribution on in-domain and cross-domain test sets. }
    \label{fig:uctta}
    \vspace{-5mm}
\end{figure}

Furthermore, the uncertainty distributions of Exp. 12 shown in Fig.~\ref{fig:uctta} exhibit a noticeable mismatch between the cross-domain CNCeleb dataset and the in-domain VoxCeleb1 dataset, providing empirical evidence that uncertainty estimation is sensitive to channel variations and recording conditions. By leveraging distribution alignment, UCDA reduces this domain mismatch, leading to a distribution in CNCeleb that is more consistent with the VoxCeleb prior, and thereby improving the reliability of uncertainty calibration.

It is worth noting that the discrepancy is not completely eliminated after adaptation. As shown in Table~\ref{tab:UCTTA}, using a smaller learning rate yields more stable and consistent improvements, whereas larger learning rates tend to degrade overall performance. This suggests that slight updates are more suitable for uncertainty calibration under domain shift, where overly aggressive adaptation may destabilize the uncertainty estimation process.

\section{Conclusion}
In this work, we proposed two complementary strategies to improve the reliability and robustness of speaker embeddings under uncertainty-aware modeling. First, we introduced an Inter- and Intra-Speaker-Aware Uncertainty Softmax that models both inter-speaker separability and intra-speaker variability, improving uncertainty estimation and speaker representations. Second, we proposed an Uncertainty-Calibrated Domain Adaptation (UCDA) framework to mitigate domain mismatch by aligning test-domain uncertainty distributions with a source-domain prior. Experiments on both in-domain and cross-domain benchmarks show consistent gains in uncertainty reliability and speaker verification performance, highlighting the importance of well-calibrated uncertainty estimation for robust speaker recognition.

\section{Acknowledgments}

The authors used generative AI to polish the language and readability of this manuscript and to assist with LaTeX formatting. All AI-generated content was reviewed by the authors, who bear full responsibility for the final work.

\bibliographystyle{IEEEtran}
\bibliography{Odyssey2026_BibEntries}
\end{document}